# Exploring the structure of time-correlated model errors in the ECMWF Data Assimilation System

Massimo Bonavita

*ECMWF, Reading, UK*

**ABSTRACT**

Model errors are increasingly seen as a fundamental performance limiter in both Numerical Weather Prediction and Climate Prediction simulations run with state of the art Earth system digital twins. This has motivated recent efforts aimed at estimating and correcting the systematic, predictable components of model error in a consistent data assimilation framework. While encouraging results have been obtained with a careful examination of the spatial aspects of the model error estimates, less attention has been devoted to the time correlation aspects of model errors and their impact on the assimilation cycle. In this work we employ a Lagged Analysis Increment Covariance (LAIG) diagnostic to gain insight in the temporal evolution of systematic model errors in the ECMWF operational data assimilation system, evaluate the effectiveness of the current weak constraint 4DVar algorithm in reducing these types of errors and, based on these findings, start exploring new ideas for the development of model error estimation and correction strategies in data assimilation.

## 1. Introduction

In Earth system prediction forecast errors grow as a result of errors in the initial conditions and model deficiencies. Model deficiencies arise from a number of concomitant factors (e.g., insufficient resolution, numerical discretisation errors, missing or approximate physical processes, etc.) which typically lead to systematic, time-correlated deviations of the model evolution from the truth. While initial condition uncertainties can be effectively characterised through ensemble data assimilation and forecasting techniques (e.g., Bonavita et al., 2012, Leutbecher et al., 2017), the treatment of systematic, time-correlated model errors has received considerably less attention, at least in the Numerical Weather Prediction (NWP) community, even though it increasingly appears to be one of the main obstacles to improved analysis and forecast accuracy and reliability.

In a data assimilation framework, Daley (1992) used previous results from Kalman Filter theory (Kailath, 1968) to show that the innovation departures of an optimal, linear DA system are not serially correlated. For a sub-optimal system, the amplitude of the serial correlations of the innovation sequence is proportional to the non-optimality of the system. Dee (1983) suggested using lagged innovation covariances in order to adaptively correct the observation and model error statistics used in a Kalman Filter based assimilation system. While the zero-lag innovation covariances are a popular tool for assessing analysis optimality in a data assimilation system (Hollingsworth and Lönnberg, 1986), the use of the lagged innovation covariances to evaluate misspecifications in model error and observation error statistics, introduced in the meteorological literature by Daley (1992a,b) has gained less traction. In Daley (1992a,b) the misspecifications were limited to the second order moments of the model and observation errors or other limitations of the assimilation system (e.g., data selection), while assuming error statistics to be white in time. In the case of time-correlated model or observation errors, lagged innovation covariances will be present even with optimally-specified error covariances. Dee and Da Silva (1998) and Dee and Todling (2000) developed and applied a Kalman Filter based framework for dealing with forecast biases evident from zero-lag innovation statistics (e.g. in the innovations from humidity radiosonde measurements). They confirmed the efficacy of their methodology by showing, among other diagnostics, how the resulting spectra of the innovations' sequence were significantly whiter than those obtained by non bias-aware analysis updates. Similar ideas have also been applied in an Ocean data assimilation context, e.g. Balmaseda et al. (2007).

The methods discussed above looked at the problem of characterising systematic model errors in observation space (i.e., they are directly driven by innovations). This may potentially limit their effectiveness in sparsely observed regions. Another possible approach is the use of analyses to infer model systematic errors. This idea has a long history in the meteorological literature, e.g. Leith (1978), and has been more recently revived in both simplified and operational analysis and modelling frameworks (e.g., Bonavita and Laloyaux, 2020, Crawford et al., 2020, Piccolo and Cullen, 2015, Mitchell and Carrassi, 2015, Carrassi and Vannitsem, 2010). These more recent works are typically based on an offline model error estimation framework: a database of analysis increments (Analysis-Background fields) is used to estimate tendency corrections that can be used during the cycling of the assimilation system or in free forecast mode (note that while the estimation procedure is offline, the estimated corrections are usually flow and/or time-dependent). It is also possible to estimate the systematic components of model error inside the data assimilation system, i.e. online estimation, based on the general technique of state augmentation (Jazwinski, 1970), where some form of explicit or parameterised model error estimate is added to the state estimate to form the control vector of the assimilation algorithm. This approach has been used in the treatment of model error in data assimilation in both the ensemble Kalman Filter context (e.g., Li et al., 2009; Zupanski and Zupanski, 2006) and in the variational framework (Derber, 1989; Zupanski, 1997; Vidard et al., 2004; Tremolet, 2006; Carrassi and Vannitsem, 2010; Laloyaux et al., 2020a), where it is generally known as weak constraint 4DVar (WC-4DVar).

As noted in Dee and Da Silva (1998), the estimation and correction of the model systematic errors is a data assimilation problem on its own which requires unbiased (or debiased) observations and a model for the evolution of the bias. A general expression for the discrete time evolution of the model error was considered in Griffith and Nichols (2001):

$\eta_{n+1} = g(x_n, \eta_n)$ (1),

where the evolution of the systematic model error component $\eta_n$ is assumed to depend on the state and thus to be correlated in time. In practice not much is known a-priori about the form of the function g and whatever previous knowledge is available should be reflected in its formulation. As discussed in Griffith and Nichols (2001), various choices are possible: persistence (i.e., $\eta_{n+1} = \eta_n$), linear evolution ($\eta_{n+1} = \mathbf{F}\eta_n$), spectral representation of the time evolution of model error (e.g., to represent diurnal error sources), linearly growing/decaying error, etc.

In the current version of weak constraint 4DVar operational at ECMWF (Laloyaux et al., 2020a) the model error is persisted from one assimilation cycle to the next, and it is only applied in the stratosphere. As already noted in Dee and Todling (2000), using persistence to predict the systematic errors implies that these estimates can only capture the component of model error which varies slowly on the timescale of the assimilation window (currently 12 hours at ECMWF). This assumption, together with the attendant choice to restrict the model error estimates to broad spatial scales (800-1000 km; Laloyaux et al., 2020a), is reasonable for the stratosphere and it is one of the reasons of the success of current WC-4DVar in dealing with stratospheric temperature biases. The situation is however different in the troposphere, due to the fact that systematic errors are smaller in size, typically driven by surface forcings and intermittent in nature. Additionally, the observational coverage is less homogeneous than in the stratosphere (Laloyaux et al., 2020b), which is bound to make the estimation problem more difficult.

In this work we first derive an expression for the time-lagged covariances of the analysis increments in the presence of systematic, time-correlated model errors (Sec. 2). A simplified version of this diagnostic is applied to the operational ECMWF data assimilation system (Sec. 3), leading to insights in the evolution of the underlying model errors. These intuitions have led to the idea of testing different options of cycling the model error estimates produced by WC-4DVar. Diagnostics of these alternative model error prognostic functions are presented in Sec. 4, while their impact on standard data assimilation and forecast performance measures is discussed in Sec. 5. Further discussion of the results and their implications for the developments of an effective strategy to account for model errors in data assimilation and forecast is provided in Sec. 6.

## 2. A Model Error Diagnostic: The Lagged Analysis Increments Covariance

In this section we review some of the effects of introducing a serially correlated model error term in a data assimilation system based on the Kalman Filter framework. This derivation is loosely based on the original work by Daley (1992a,b), and references therein.

In a system governed by linear dynamics the evolution of the forecast error from one assimilation time (t=n) to the next (t=n+1) can be written as:

$$\varepsilon_{n+1}^f = \mathbf{M}\varepsilon_n^a + \varepsilon_n + \eta_n \qquad (2).$$

In Equation (2) $\mathbf{M}$ is the prognostic model used to cycle the assimilation system, $\varepsilon_n$ is an unbiased ($\langle \varepsilon_n \rangle = 0$), time uncorrelated ($\langle (\varepsilon_{n+1}^q)^T \varepsilon_n^q \rangle = 0$), stochastic noise term completely characterised by its covariance matrix ($\mathbf{Q}_s = \langle \varepsilon_n(\varepsilon_n)^T \rangle$). Differently from typical Kalman Filter applications in the atmospheric sciences, we consider here an additional model error term $\eta_n$ which we allow to be biased ($\langle \eta_n \rangle \neq 0$) and serially correlated ($\langle (\eta_{n+1})^T \eta_n \rangle \neq 0$), but still uncorrelated with the stochastic noise term $\varepsilon_n^q$ and the observational errors.

A linear, unbiased assimilation update in the presence of a new batch of observations at time t=n+1 will produce an analysis whose errors can be characterised as:

$$\varepsilon_{n+1}^a = (\mathbf{I} - \mathbf{K}_{n+1}\mathbf{H})\varepsilon_{n+1}^f + \mathbf{K}_{n+1}\varepsilon_{n+1}^o \qquad (3),$$

where $\mathbf{H}$ is the observation operator of the observing system (which we assume stationary for simplicity), and $\mathbf{K}_{n+1}$ is a generic gain matrix also valid at time t=n+1.

In this theoretical framework, Daley, 1992b, rederived the classical result that the lagged innovation covariance $\mathbf{C}_m^n \stackrel{\text{def}}{=} \langle (y_m - \mathbf{H}x_m^f)(y_n - \mathbf{H}x_n^f)^T \rangle = 0$ for $m \neq n$ if the gain matrix is optimal (Kalman gain) and if model and observation errors are neither mutually nor serially correlated. The whiteness of the time series of innovation departures for specific observing systems can thus be interpreted as a diagnostic of optimality of the assimilation system under the stated assumptions on model and observation errors.

We are now interested to see how the lagged innovation result transfers to the case where we are interested in analysing the lagged time covariance of the analysis increments, i.e.:

$$\mathbf{A}_m^n \stackrel{\text{def}}{=} \langle (x_m^a - x_m^f)(x_n^a - x_n^f)^T \rangle = \langle \Delta x_m^a (\Delta x_n^a)^T \rangle \qquad (4).$$

In particular, we want to see the form of the lag-1 covariance of the analysis increments for a DA system which evolves under Equations 1 and 2, i.e., $\mathbf{A}_{n+1}^n \stackrel{\text{def}}{=} \langle (x_{n+1}^a - x_{n+1}^f)(x_n^a - x_n^f)^T \rangle$

Using Equations (1) and (2), and making the standard assumption of statistical independence between observations and background and analysis errors at different update times, we can write:

$$\mathbf{A}_{n+1}^n \stackrel{\text{def}}{=} \langle (x_{n+1}^a - x_{n+1}^f)(x_n^a - x_n^f)^T \rangle = \langle \mathbf{K}_{n+1}\left(\varepsilon_{n+1}^o - \mathbf{H}(\mathbf{M}\varepsilon_n^a + \varepsilon_n^q + \eta_n)\right)(\varepsilon_n^a - \varepsilon_n^f)^T \rangle =$$

$$= -\mathbf{K}_{n+1}\mathbf{HM}\,\langle \varepsilon_n^a (\varepsilon_n^a - \varepsilon_n^f)^T \rangle - \mathbf{K}_{n+1}\mathbf{H}\,\langle \eta_n (\varepsilon_n^a - \varepsilon_n^f)^T \rangle \qquad (5).$$

The first term in Equation (5) can be expanded as:

$$-\mathbf{K}_{n+1}\mathbf{HM}\,\langle \varepsilon_n^a (\varepsilon_n^a - \varepsilon_n^f)^T \rangle$$

$$= -\mathbf{K}_{n+1}\mathbf{HM}\left(\langle \varepsilon_n^a (\varepsilon_n^a)^T \rangle - \left((\mathbf{I} - \mathbf{K}_n\mathbf{H}_n)\,\langle \varepsilon_n^f (\varepsilon_n^f)^T \rangle + \mathbf{K}_{n+1}\,\langle \varepsilon_n^o (\varepsilon_n^f)^T \rangle\right)\right) =$$

$$= -\mathbf{K}_{n+1}\mathbf{HM}\left(\mathbf{P}_n^a - (\mathbf{I} - \mathbf{K}_n\mathbf{H}_n)\mathbf{P}_n^f\right) \qquad (6).$$

The right hand size of Equation (6) vanishes in the case of a DA system using an optimal gain matrix (a Kalman gain), i.e. a matrix that minimises the expected analysis error covariance: this is in fact equivalent to the requirement that the residual analysis error be orthogonal to the analysis increment. In a DA system where the model is not affected by a serially correlated model bias we thus recover a similar property of whiteness of the time series of analysis increments as that valid for the lagged innovation covariances.

In the case of time-correlated model error, however, the second term of Equation (5) can still give a significant contribution to lag-1 analysis increment covariance. Assuming this term is the dominant contribution (i.e., optimal linear analysis update), we get through repeated applications of Eq. 2, 3:

$$\mathbf{A}_{n+1}^n \cong -\mathbf{K}_{n+1}\mathbf{H}\,\langle \eta_n (\varepsilon_n^a - \varepsilon_n^f)^T \rangle = \mathbf{K}_{n+1}\mathbf{H}\,\langle \eta_n (\varepsilon_n^f)^T \rangle \mathbf{H}^T \mathbf{K}_n^T$$

$$= \mathbf{K}_{n+1}\mathbf{H}\langle \eta_n (\eta_{n-1})^T \rangle \mathbf{H}^T \mathbf{K}_n^T + \mathbf{K}_{n+1}\mathbf{H}\langle \eta_n (\varepsilon_{n-1}^a)^T \rangle \mathbf{M}^T \mathbf{H}^T \mathbf{K}_n^T$$

$$= \mathbf{K}_{n+1}\mathbf{H}\langle \eta_n(\eta_{n-1})^T\rangle \mathbf{H}^T\mathbf{K}_n^T + \mathbf{K}_{n+1}\mathbf{H}\langle \eta_n(\eta_{n-2})^T\rangle(\mathbf{I} - \mathbf{K}_{n-1}\mathbf{H})^T\mathbf{M}^T\mathbf{H}^T\mathbf{K}_n^T$$

$$+ \mathbf{K}_{n+1}\mathbf{H}\langle \eta_n(\varepsilon_{n-2}^a)^T\rangle(\mathbf{I} - \mathbf{K}_{n-1}\mathbf{H})^T\mathbf{M}^T\mathbf{H}^T\mathbf{K}_n^T$$

$$= \mathbf{K}_{n+1}\mathbf{H}\langle \eta_n(\eta_{n-1})^T\rangle\mathbf{H}^T\mathbf{K}_n^T + .. + \mathbf{K}_{n+1}\mathbf{H}\langle \eta_n(\eta_{n-k})^T\rangle((\mathbf{I} - \mathbf{K}_{n-k+1}\mathbf{H})^T\mathbf{M}^T .. (\mathbf{I} -$$

$$\mathbf{K}_{n-1}\mathbf{H})^T\mathbf{M}^T)\mathbf{H}^T\mathbf{K}_n^T + .. \qquad (7).$$

The lag-1 analysis increment covariance is thus given by the sum of a leading order contribution proportional to the lag-1 model error covariance and higher order contributions of the lagged model error covariances at increasing time separations. The higher order contributions are expected to taper off rapidly, not only because of the decreasing covariances at increasing lag intervals, but also due to the repeated application of contraction operators in the analysis updates. The other important, and in a sense obvious, aspect is that the information we obtain on the model error through the analysis increments is always mediated by the observing system (through the application of the observation operator $\mathbf{H}$) and by the projection of this information back into model space through the Kalman gain matrix $\mathbf{K}$. In other words, all the information we have on model error depends on the distribution and quality of available observations and the optimality of the analysis update.

The analysis increment lagged covariances for higher lag times have more complex analytical expressions. To give an example, the lag-2 covariance expansion to leading terms is:

$$\mathbf{A}_{n+2}^n \stackrel{\text{def}}{=} \langle (x_{n+2}^a - x_{n+2}^f)(x_{n+2}^a - x_n^f)^T\rangle \cong \mathbf{K}_{n+2}\mathbf{H}\langle \eta_{n+1}(\eta_{n-1})^T\rangle\mathbf{H}^T\mathbf{K}_n^T + \mathbf{K}_{n+2}\mathbf{H}\mathbf{M}(\mathbf{I} -$$

$$\mathbf{K}_{n+1}\mathbf{H})\langle \eta_n(\eta_{n-1})^T\rangle\mathbf{H}^T\mathbf{K}_n^T + ... \qquad (8).$$

By induction, the k-lagged covariance:

$$\mathbf{A}_{n+k}^n \stackrel{\text{def}}{=} \langle (x_{n+k}^a - x_{n+k}^f)(x_{n+2}^a - x_n^f)^T\rangle \cong \mathbf{K}_{n+k}\mathbf{H}\langle \eta_{n+k-1}(\eta_{n-1})^T\rangle\mathbf{H}^T\mathbf{K}_n^T + \mathbf{K}_{n+k}\mathbf{H}\mathbf{M}(\mathbf{I} -$$

$$\mathbf{K}_{n+k-1}\mathbf{H})\langle \eta_{n+k-2}(\eta_{n-1})^T\rangle\mathbf{H}^T\mathbf{K}_n^T + ... \qquad (9),$$

As in the lag-1 covariance, the lag-k covariance is given by a term proportional to the corresponding lag-k model error covariance plus a summation over evolved model error covariances at increasing lag times. Again, the contributions of these terms are expected to decrease rapidly with increasing lag time.

To go further in this analysis would require making explicit assumptions about the form of the random process governing the evolution of the $\eta_n$ model error term. However, the form of Equations (7-9) already suggests that lagged covariances of analysis increments are informative about the structure of model errors which have non-negligible time correlations over scales longer than the length of one assimilation window. We will explore them empirically in the following sections.

## 3. Model error diagnostics using lagged analysis increments covariances

We start here by considering the structure of the lagged analysis increment covariances in cycling assimilation experiments based on the operational version of the IFS at the time of writing (cycle 47R1, July 2020). The k-lagged covariance for a timeseries of length n with overall mean μ and overall variance σ² has the standard expression:

$$R_k = \frac{1}{n}\sum_{i=1}^{n-k} \frac{(y_i-\mu)(y_{i+k}-\mu)}{\sigma^2} \qquad (10).$$

The timeseries of analysis increments considered in this work showed no significant trend in the mean and in the variance, so it is justified to assume weak stationarity conditions and thus that covariances depend only on the lag k. The analysis increments are valid at the start of the 12-hour assimilation window, which in the ECMWF analysis cycle are at 09 and 21 UTC. We start by considering the time-lagged covariances as a function of time lag (this is also known as auto-correlogram in time series analysis) of the analysis increments for the main 4D-Var control variables collected over one month (August 2019). In Figure 1 we plot the global averages of these quantities for different model levels for a standard IFS assimilation cycle at operational resolution (T1279 spectral truncation, ~10km grid point spacing) using strong constraint 4D-Var in the analysis update (i.e., model error is neither estimated nor corrected for in this setup). One common feature of the plots in Figure 1 is the presence of a significant diurnal cycle of the analysis increments, which is highlighted by the 24 hour periodicity in their autocovariances and is superimposed on the expected uniform decorrelation trend of the analysis increments over longer timescales. The size of the diurnal cycle signal is largest near the surface (model level 137) and for integrated variables like surface pressure (SP), and decreases with height, but is still visible throughout the model column. Another feature is that the size of the autocovariances is largest for the mass fields (temperature and SP) and humidity, while it is relatively small (less than 10%) for wind variables.

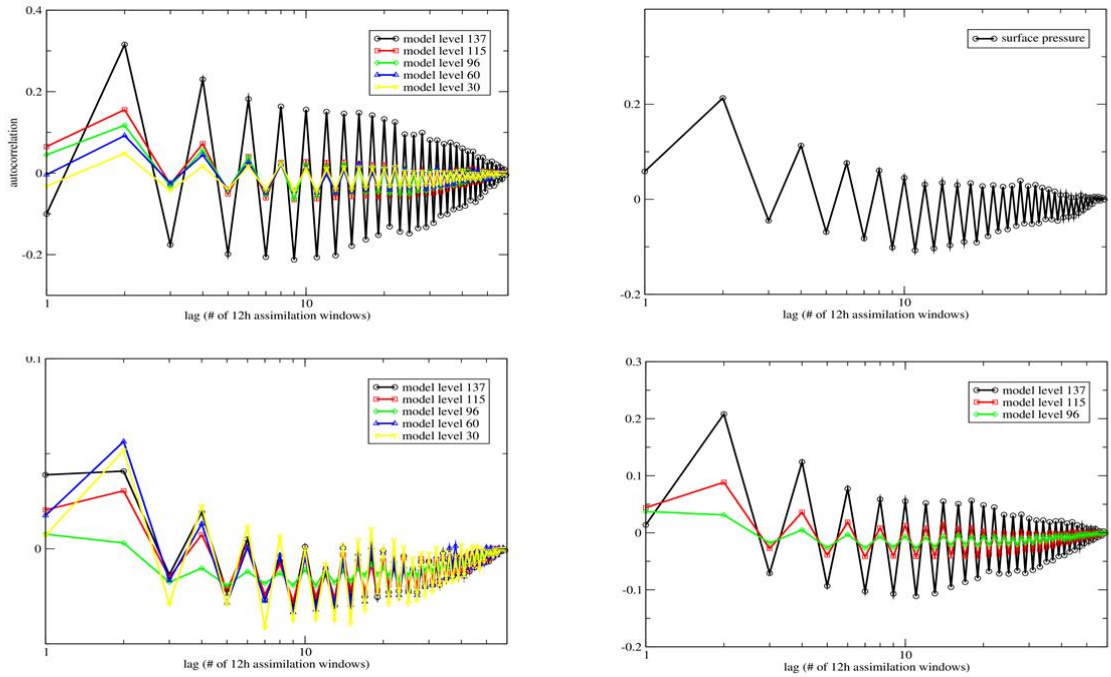

Figure 1: *Lagged analysis increment covariances from a strong constraint 4D-Var experiment for temperature (top left panel), surface pressure (top right), vorticity (bottom left) and specific humidity (bottom right). The covariances have been computed using 60 samples of analysis increments collected over August 2019. (Model levels: 137=surface layer (black); 115~850 hPa (red); 96~500 hPa (green); 60~100 hPa (blue); 30~10 hPa (yellow)).*

Another significant aspect of the analysis increments autocovariances is that the relatively small values of the globally averaged lag-1 autocovariances have different origins near the surface and in the lowest model levels with respect to the free atmosphere. The lag-1 autocovariances at the surface are still large in size but their global average is small due to the presence of large areas of compensating positive and negative correlations (Figure 2, top row). Above the boundary layer, autocovariances tend, on the other hand, to become smaller in absolute value and noisier, and retain discernible spatial structures only in specific areas ((Figure 2, bottom row).

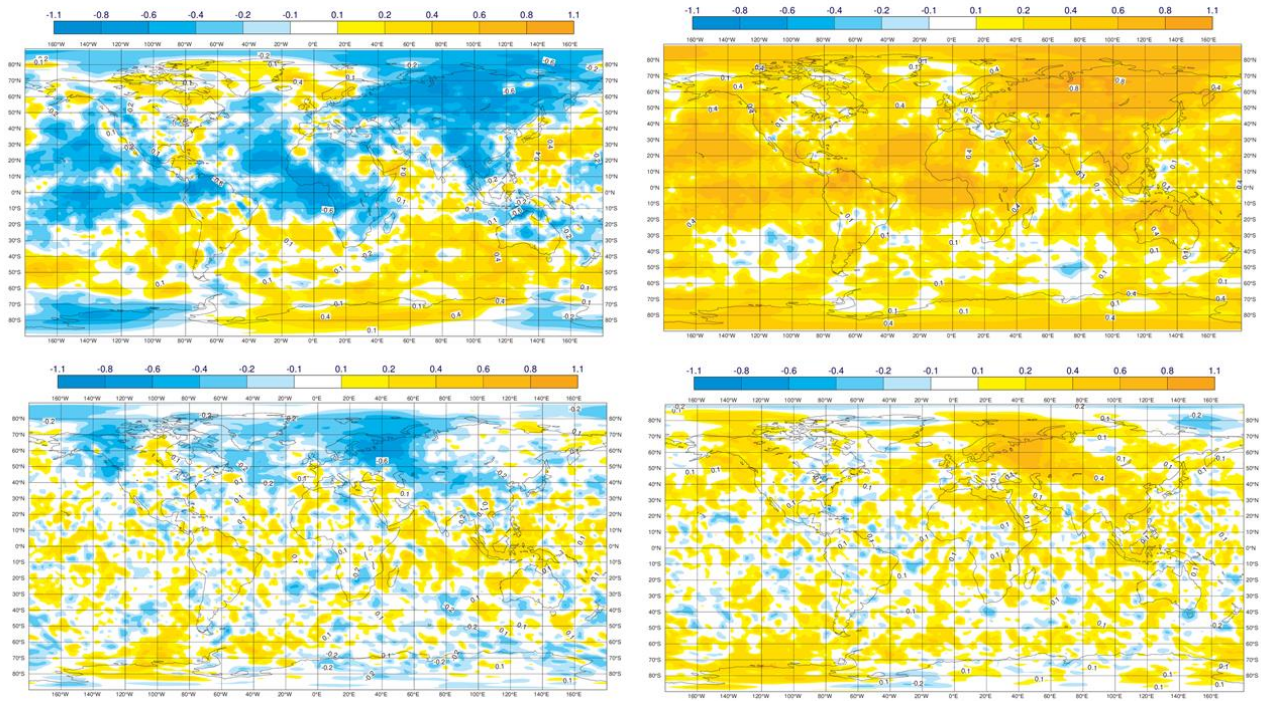

Figure 2: *Top row: Lag-1 (left) and lag-2 (right) autocovariance of the temperature analysis increments near the surface (model level 137). Bottom row: As top row for model level 60 (~100 hPa).*

To give further appreciation of the spatial structure of the autocovariances, we compare in Figure 3 the lag-1 and lag-2 autocovariances for surface pressure analysis increments (top row) and the average analysis increments for the 00UTC and 12UTC analysis updates. The lag-1 SP autocovariances present large areas of positive correlations but also visible regions of anti-correlations, e.g. over most of the African continent, south-east Asia, ITCZ, west Pacific, while the lag-2 autocovariance presents positive correlations everywhere. The regions of negative values in the lag-1 autocovariances correlate well with the regions where the average analysis increments change sign in the 00 and 12 UTC analyses. This confirms that the oscillatory behaviour seen in the autocovariance plots is largely driven by diurnal changes in the average analysis increments which, in the theoretical framework presented in Sec. 2, derive from a diurnal mode of model error.

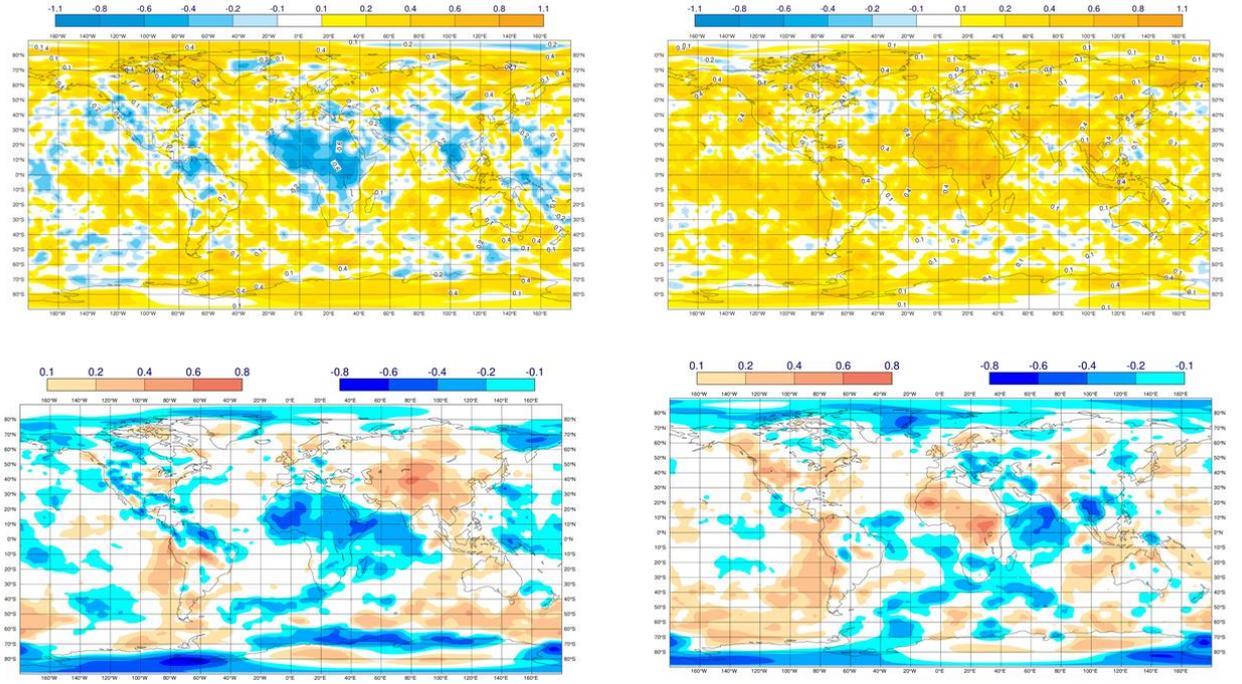

Figure 3: *Top row: Lag-1 (left) and lag-2 (right) autocovariance of the surface pressure analysis increments. Bottom row: mean surface pressure analysis increments for the 00UTC analysis update (left) and 12UTC analysis update (right).*

## 4. Estimating model error with weak constraint 4DVar

The current ECMWF implementation of Weak Constraint 4D-Var (WC-4DVar, Laloyaux et al., 2020a, b) works by estimating, together with the state of the atmosphere at the start of the assimilation window ($x_0$), a stationary model error tendency $\eta$ over the 12 hour assimilation window:

$$J_{WC}(x_0, \eta) = \frac{1}{2}(x_0 - x_0^b)^T \mathbf{B}^{-1}(x_0 - x_0^b) + \frac{1}{2}\sum_{k=0}^{N}(H(x_k) - y_k)^T \mathbf{R}_k^{-1}(H(x_k) - y_k) +$$

$$\frac{1}{2}(\eta - \eta^b)^T \mathbf{Q}^{-1}(\eta - \eta^b) \qquad (10)$$

In the current operational configuration (IFS Cycle 47r1, July 2020) the model error estimation and correction is only performed in the stratosphere above 100 hPa, and the model error estimates are cycled from one analysis update to the next through the identity operator (i.e., the analysed model error at time t=n-1 is used as background term for the WC-4DVar update at time t=n). Based on the diagnostics presented in Figure 1, this appears to be a reasonable choice for the estimation of model error in the stratosphere, less so in the troposphere. To better understand the connections between the analysis increment covariance diagnostics and the model error estimates produced by WC-4DVar, we have run three WC-4DVar experiments with different model error cycling configurations and in which the model error corrections are active throughout the model column. Note that in these experiments we have used the current (July 2020) operational model error covariance matrix, which is likely to be sub-optimal in the troposphere, for the reasons discussed in Laloyaux et al., 2020a, 2020b. Thus, the results presented below should be interpreted as sensitivities of WC-4DVar to different model error cycling strategies and not as an absolute measure of current or potential WC-4DVar performance. The main characteristics of all the experiments discussed in this work are summarised in Table 1.

|  | SC-4DVar | Operational WC-4DVar | Restarted WC-4DVar | Cycled WC-4DVar | Diurn Cycled WC-4DVar |
|---|---|---|---|---|---|
| *Resolution/Vertical Levels* | TCo1279/L137 | TCo1279/L137 | TCo1279/L137 | TCo1279/L137 | TCo1279/L137 |
| *Model error* | No | Yes | Yes | Yes | Yes |
| *Model error active layer* | N.A. | Levels 1-60 | Levels 1-137 | Levels 1-137 | Levels 1-137 |
| *Model error background ($\boldsymbol{\eta}^b$)* | N.A. | $\boldsymbol{\eta}^b = \boldsymbol{\eta}^a_{n-1}$ | $\boldsymbol{\eta}^b = \boldsymbol{0}$ | $\boldsymbol{\eta}^b = \boldsymbol{\eta}^a_{n-1}$ | $\boldsymbol{\eta}^b = \boldsymbol{\eta}^a_{n-2}$ |

Table 1: *Main characteristics of the assimilation experiments discussed in the text.*

**4.1 Restarted WC-4DVar**

In this configuration of WC-4DVar the model error estimates are not cycled from one assimilation update to the next, but set to zero at the start of each 4D-Var update (i.e., $\boldsymbol{\eta}^b = \boldsymbol{0}$ in Equation (10)). As a consequence, only observations available in each 12-hour assimilation window can affect the model error estimates. The motivation for this experiment is to assess how much of the perceived model error can be recovered by WC-4DVar using only 12 hours of observational information. This will also provide us with a baseline to evaluate the effectiveness of the model error cycling configurations described below.

As shown in Figure 4, the introduction of a model error term in 4DVar produces, on average, a relatively homogeneous reduction in the size of the analysis increments of around 2-3%. This reduction is expected, as the introduction of a model error term provides additional degrees of freedom for 4DVar to reduce the observations' misfits beyond the corrections to the initial state.

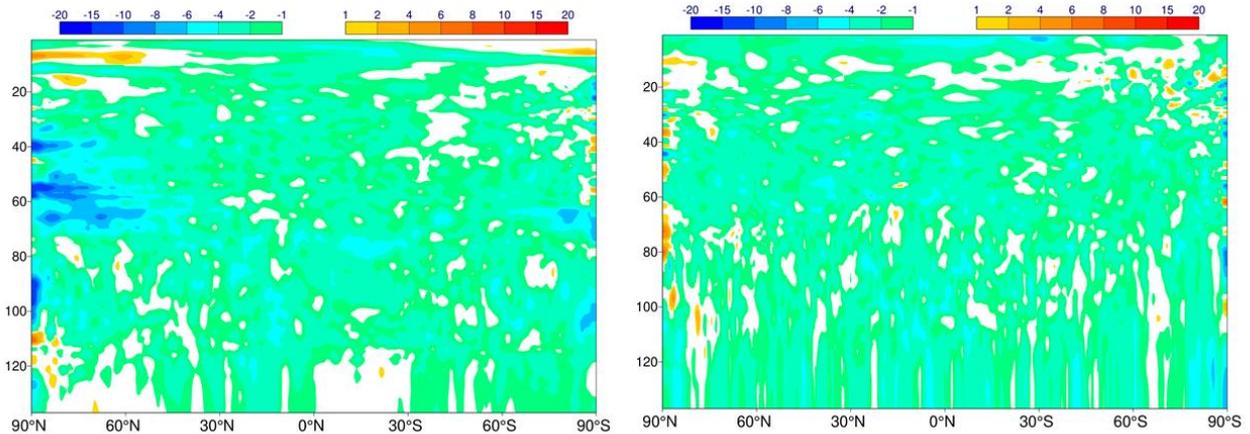

Figure 4: *Relative difference in the standard deviation of the temperature (left) and vorticity (right) analysis increments of the restarted WC-4DVar experiment and the strong constraint 4DVar experiment). The statistics have been computed using 60 samples of analysis increments collected over August 2019. (Model levels legend: 137=surface layer; 120~900 hPa; 100~600 hPa; 60~100 hPa; 40~30 hPa; 20~3 hPa).*

The restarted WC-4DVar experiment also shows the expected reduction in the mean analysis increments for both the 00 and 12UTC analysis, but this effect is not large and mainly confined to the free troposphere (Figure 5). This is confirmed by the fact that the estimated model error tendencies have significant magnitudes only in the 900-100 hPa vertical layer (model level 120 to 60; Figure 6). Interestingly, the variability of the model error estimates (Figure 6, right panel) is comparable to the absolute value of their mean (Figure 6, left panel). This suggests that in the restarted WC-4DVar configuration the impact of the errors of the day (i.e., predictability errors) on the model error estimates is still significant.

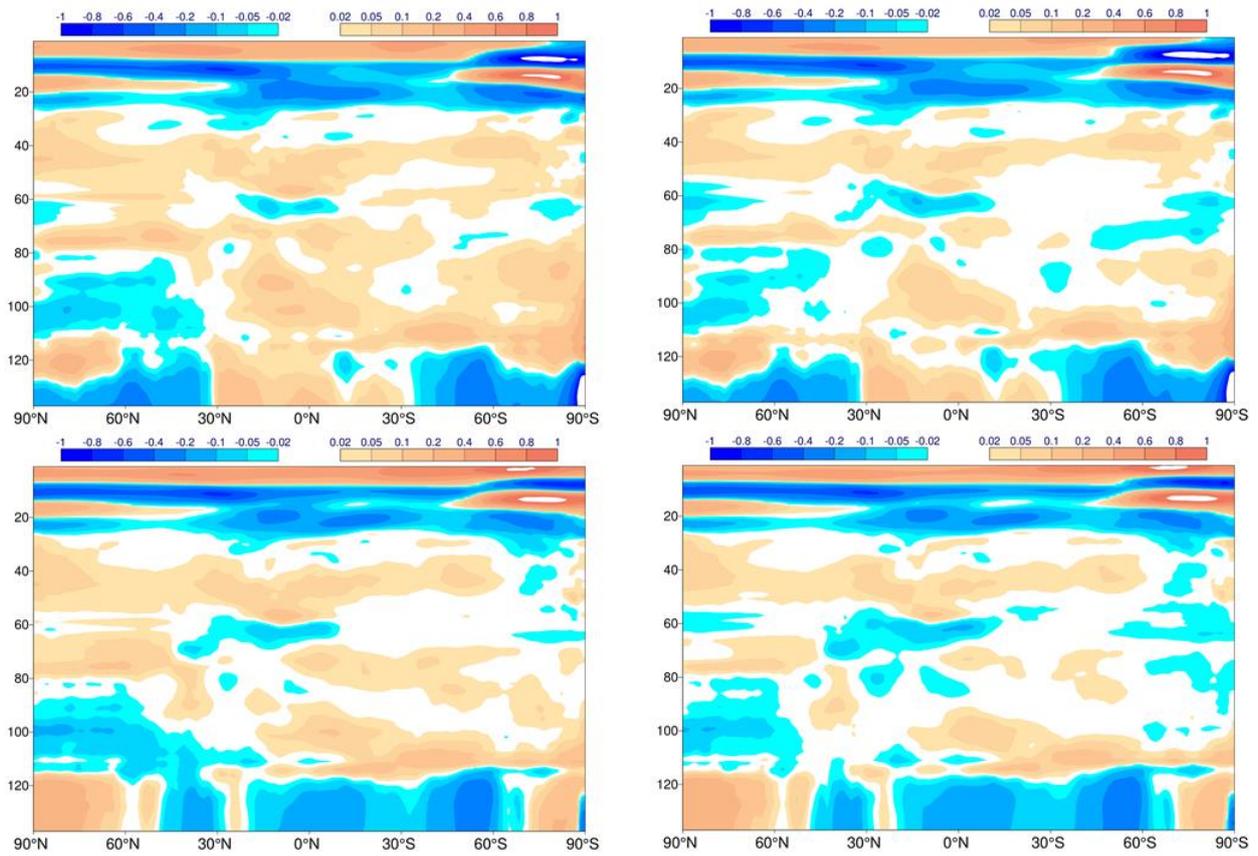

Figure 5: *Average temperature analysis increments at 00UTC (top row) and 12UTC (bottom row) for the strong constraint 4DVar experiment (left column) and the restarted weak constraint 4DVar experiment (right column). Statistics collected over the August 2019 period.*

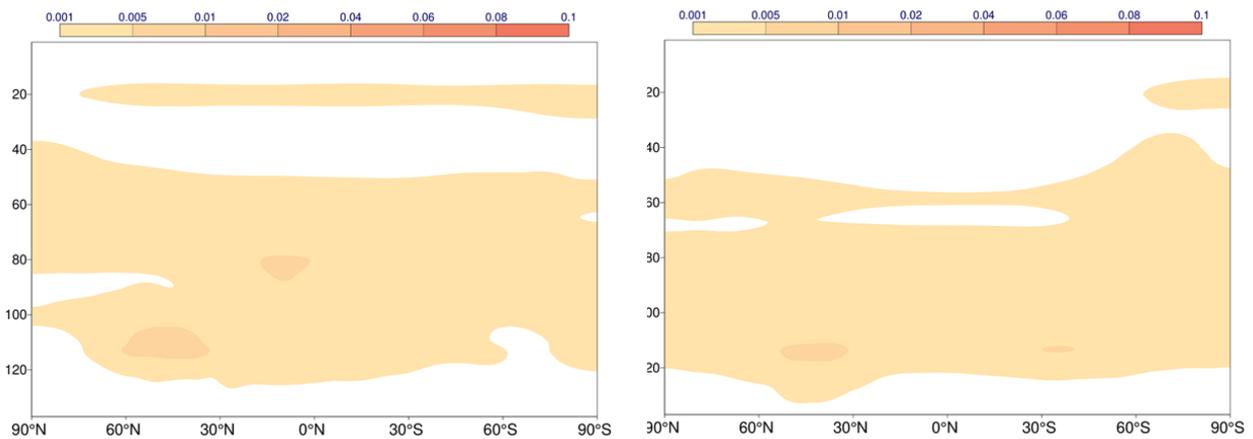

Figure 6: *Absolute value of the mean (left panel) and standard deviation (right panel) of the temperature model error tendency estimated in the restarted WC-4DVar experiment. Values averaged over the month of August 2019, Units: Kelvin/hour.*

Finally, the autocovariances of the model error estimates produced in this experiment are shown in Figure 7 (Autocovariances for humidity are not shown because humidity is not currently part of the model error control vector). Comparing Figure 7 with Figure 1, it is apparent that the structure of the estimated model error autocovariances show similarities to that of the analysis increments, e.g., in terms of the signature of the diurnal cycle and the longer range time decay, but also notable differences. One difference is that the model error autocovariances are generally larger in magnitude than the corresponding autocovariances of analysis increments. Another is that their diurnal variation is more pronounced than it is in the analysis increment autocovariances and does not appear to decrease in magnitude with height. Both these effects can be, at least partly, explained by the fact that the lagged analysis autocovariances are effectively time integrals of the model error over more than one assimilation window (Eq. 7-9). However, they also suggest that cold starting the model error background at each assimilation cycle can limit the ability of WC-4DVar to correct for model errors, for at least two different reasons. Firstly, where model errors show a significant diurnal cycle, the model error corrections learned in one assimilation window are not good predictors of model error in the following assimilation window, which can explain the increased diurnal cycle of the model error estimates. Secondly, the lagged analysis increments autocovariances indicate the presence of model errors acting on time ranges longer than the assimilation window length, which the restarted WC-4DVar cannot capture by construction.

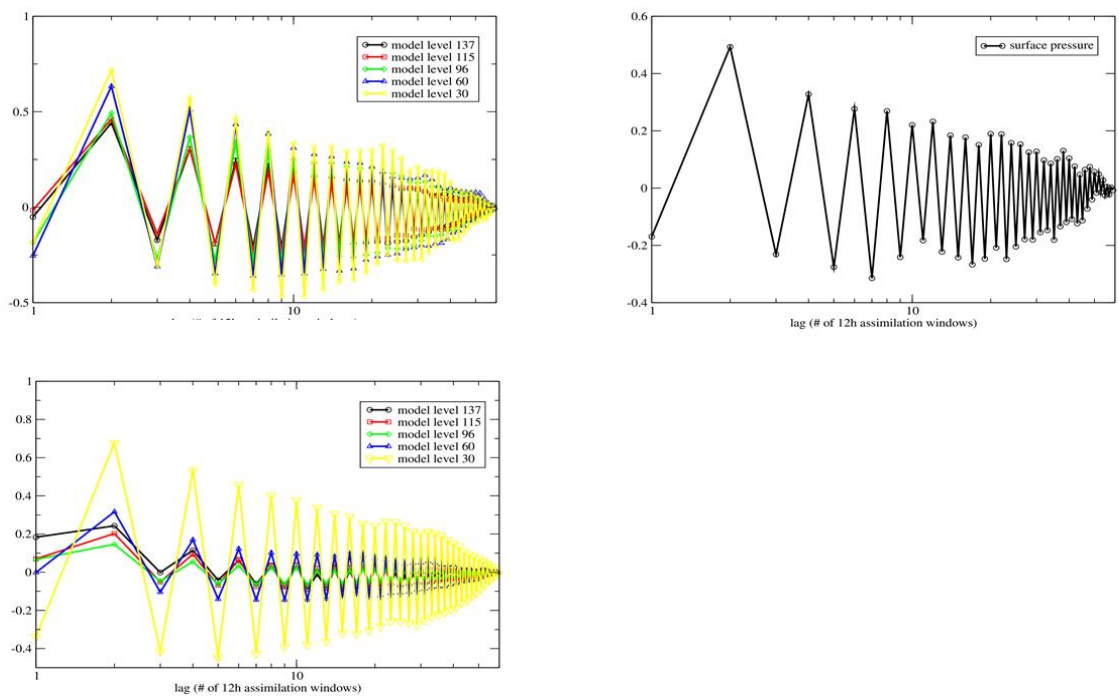

Figure 7: *Lagged model error covariances from the restarted weak constraint 4D-Var experiment for temperature (top left panel), surface pressure (top right), vorticity (bottom left)*

## 4.2 Cycled WC-4DVar

In this configuration of WC-4DVar the model error estimates are cycled from one assimilation update to the next using the identity operator, i.e. the background values for the WC-4DVar update for the current assimilation update are the model errors estimated in the previous assimilation update. This WC-4DVar configuration mirrors the current operational ECMWF WC-4DVar as described in Laloyaux et al. 2020a, except that the model errors are estimated and applied throughout the model column and not only in the stratosphere. This cycled version of WC-4DVar has a broader and stronger impact on the analysis than the restarted version. This is visible in terms of analysis increments (Figure 8, left panel), which are generally smaller, especially in the stratosphere; and in terms of the spatial distribution of the model error tendency corrections (Figure 8, right panel, to be compared with Figure 6, left panel). The right panel of Figure 8 also indicates that the size of the model error corrections is significantly increased in the cycled WC-4DVar. This suggests that, as expected, the cycled 4D-Var configuration is better able to capture model error modes which vary slowly on longer timescales. It is also noticeable that the model error corrections tend to be larger in specific geographical areas and vertical layers, e.g., tropical troposphere, top of the boundary layer, top of the model, which are known to be regions of larger systematic errors of the model.

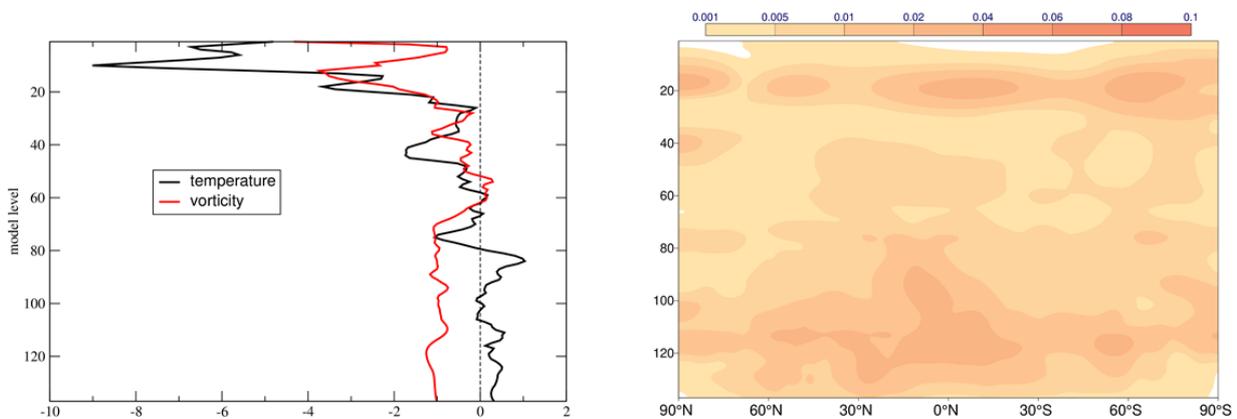

Figure 8: *Left panel: Vertical profile of the relative difference in the standard deviation of the analysis increments of the cycled WC-4DVar experiment and the restarted WC-4DVar experiment for temperature (black) and vorticity (red). Right panel: Absolute value of the mean of the temperature model error tendencies estimated in the cycled WC-4DVar experiment. Values averaged over the*

*month of August 2019, Units: Kelvin/hour.*

The profiles of the mean absolute analysis increments (Figure 9, left panel), confirm that the cycled WC-4DVar is more effective than the restarted WC-4DVar in dealing with systematic errors in the free troposphere and stratosphere. Improvements are also seen in the top 20 model levels, where the size of the increments is reduced, but not eliminated ( this is assumed to be due to the sharp vertical inversions in the sign of the systematic errors in this layer, combined with the broad vertical scales of the current model error covariance matrix). On the other hand, the cycled WC-4DVar setup does not seem to be effective in reducing systematic errors in the boundary layer, which have a significant diurnal cycle component. This is confirmed by the plot in Figure 9, right panel, where the relative difference in the absolute values of the lag-1 and lag-2 analysis increment covariances between the cycled WC-4DVar and the strong constraint 4DVar are shown (black lines in the plot). Despite some noisiness, especially towards the model top, the cycled WC-4DVar shows some reduction in the magnitude of the lagged covariances in the vertical layer between model levels 100 and 40 (approx. 600-30 hPa), which corresponds to the layer where the average analysis increments have been more clearly reduced (Figure 9, left panel), but not in the boundary layer.

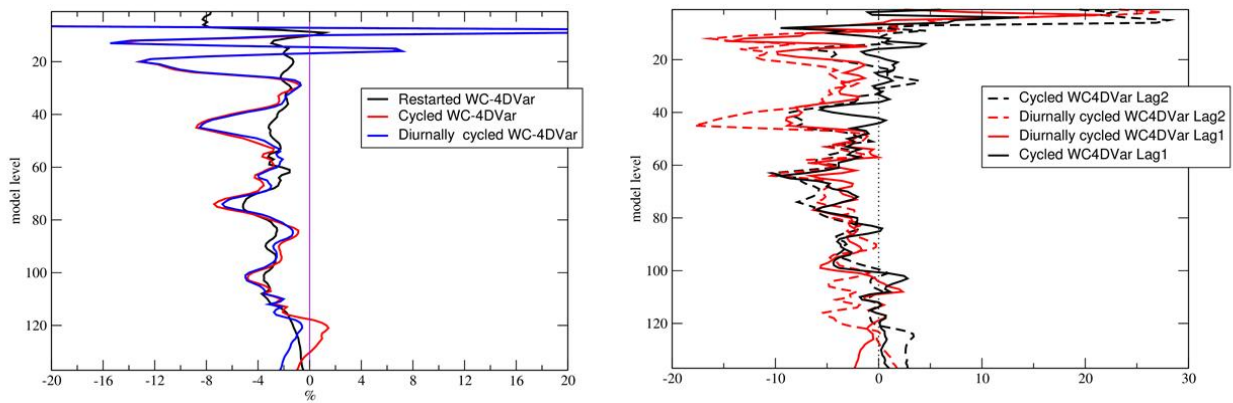

Figure 9 *Left panel: Vertical profile of the relative difference in the mean absolute values of the temperature analysis increments of the restarted WC-4DVar (black), the cycled WC-4Dvar (red) and the diurnally cycled WC-4Dvar (blue) with respect to the strong constraint 4Dvar baseline experiment. Right panel: Vertical profile of the relative difference in the absolute values of the lag-1 (continuous lines) and lag-2 (dashed lines) covariances of the analysis increments of the cycled WC-4DVar experiment (black) and diurnally cycled WC4DVar (red) with respect to the strong constraint 4DVar experiment. Values averaged over the month of August 2019.*

A final diagnostic of interest is the auto-correlogram of the model error estimates, which is shown in Figure 10. Comparing it with the corresponding plots for the restarted WC-4DVar experiment (Figure 7) and the analysis increments' lagged covariances in Figure 1, the most striking difference is how the effect of the diurnal component of model error is reduced (and almost eliminated in the boundary layer and stratosphere). This is a direct consequence of the cycling of the model error estimates which, with the model error covariance matrix used in these experiments, induces a decorrelation time of about one to two weeks in the model error estimates, depending on model level, which filters out the diurnal variability of the model error estimates.

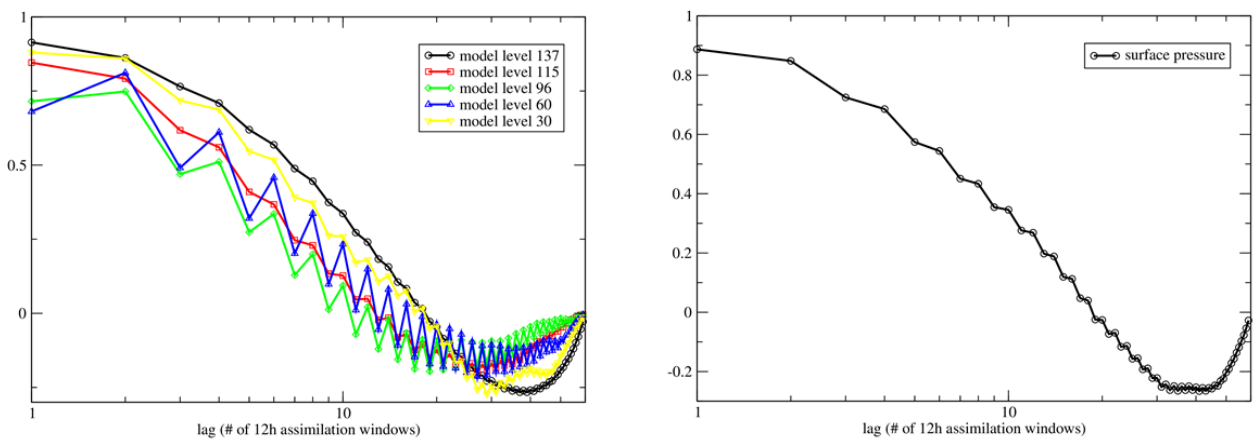

Figure 10: *Lagged model error covariances from the cycled weak constraint 4D-Var experiment for temperature (left panel) and surface pressure (right). Other details as in Figure 1.*

**4.3 Diurnally cycled WC-4DVar**

From the discussion of the two previous configurations of weak constraint 4DVar, it appears that cycling the model error estimates through persistence is useful in order to accumulate information on model error over many assimilation windows, but it also has the undesirable consequence of smoothing the diurnal component of model error that is visible in the mean analysis increment statistics. A possible solution could be to cycle the error estimates with a 24 hour period, i.e. the model error background estimate used in the time t=n analysis update is the model error analysis at t=n-2. We call this configuration "diurnally cycled" WC-4DVar.

No large scale differences are visible with respect to the cycled WC-4DVar setup are seen in either the standard deviation of the analysis increments or the magnitude and distribution of the model error corrections (not shown). However, a closer look at the profile of the mean absolute analysis increments (Figure 9, left panel) shows that in the boundary layer (model levels 120 to 137) the diurnally cycled version of WC-4DVar marginally reduces the magnitude of the mean analysis increments. In the rest of the troposphere and in the stratosphere the two cycled WC-4DVars behave similarly and both have a larger impact on mean analysis increments than the restarted WC-4DVar.

In the other measure of the effectiveness of weak constraint 4DVar to reduce the correlated components of model error in the assimilation cycle, the lagged analysis increment covariance, the diurnally cycled WC4DVar appears more effective than the cycled WC4DVar at reducing the magnitude of these covariances, particularly in the boundary layer and middle/upper stratosphere (Figure 9, right panel).

Finally, we present the evolution of the lagged covariances of the model error estimates in this WC-4DVar configuration in Figure 10. As expected, the evolution of the covariances as a function of lag time is dominated by a diurnal cycle, over which a slow temporal decay is superimposed.

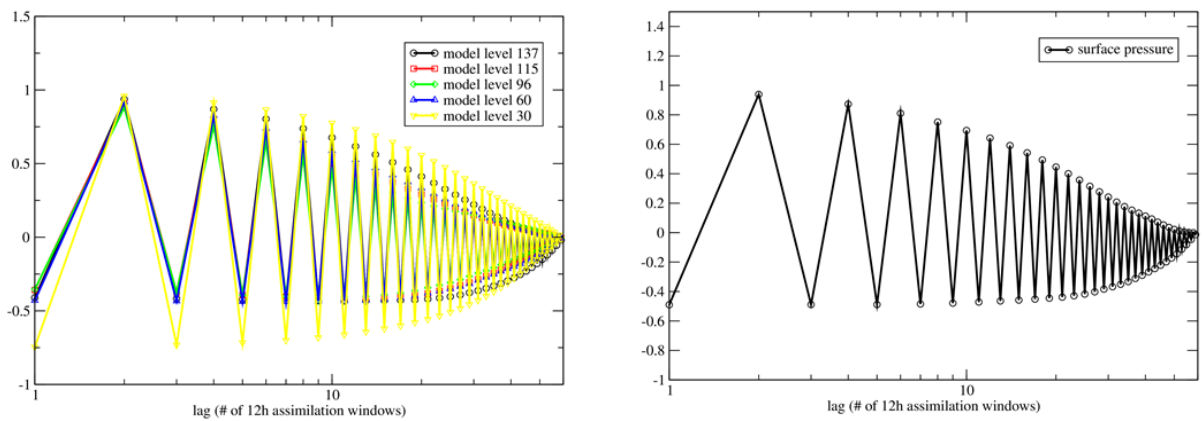

Figure 10: *Lagged model error covariances from the diurnally cycled weak constraint 4D-Var experiment for temperature (left panel) and surface pressure (right). The covariances have been computed using 60 samples of model error estimates collected over August 2019. (Model levels: 137=surface layer (black); 115~850 hPa (red); 96~500 hPa (green); 60~100 hPa (blue); 30~10 hPa (yellow)).*

One of the main assumptions underlying the current ECMWF implementation of weak constraint 4DVar is the separation of large scale, slowly evolving model errors, which can be estimated and corrected in a deterministic assimilation framework, from short scale, fast error modes, which are assumed to be zero-mean and whose second moments can only be estimated through stochastic perturbations to the prognostic model in the cycling of an Ensemble of Data Assimilations (EDA, Bonavita et al., 2012). It is thus interesting to check if and to what extent the various WC-4DVar configurations discussed here impact the length scale of the analysis increments. The results of this diagnostic are presented in Figure 11, where the left panel shows the profile of the globally averaged length scale of the temperature analysis increments for the strong constrain 4DVar experiment, while the right panel shows the relative difference in the length scales of the three WC 4DVar configurations tested here with respect to the strong constraint 4DVar reference. The length scales of the analysis increment are typically in the 200 to 300 km range, slowly increasing with height, with a small increase in the bottom model layers and a very large increase in the top 20 model levels, where the model is affected by large systematic errors connected to treatment of the sponge layer (Polichtchouk et al., 2021). All three WC 4DVar configurations manage to slightly reduce the average length scales of the increments, with the two cycled configurations having a larger impact and the diurnally cycled WC 4DVar having the largest impact in the boundary layer.

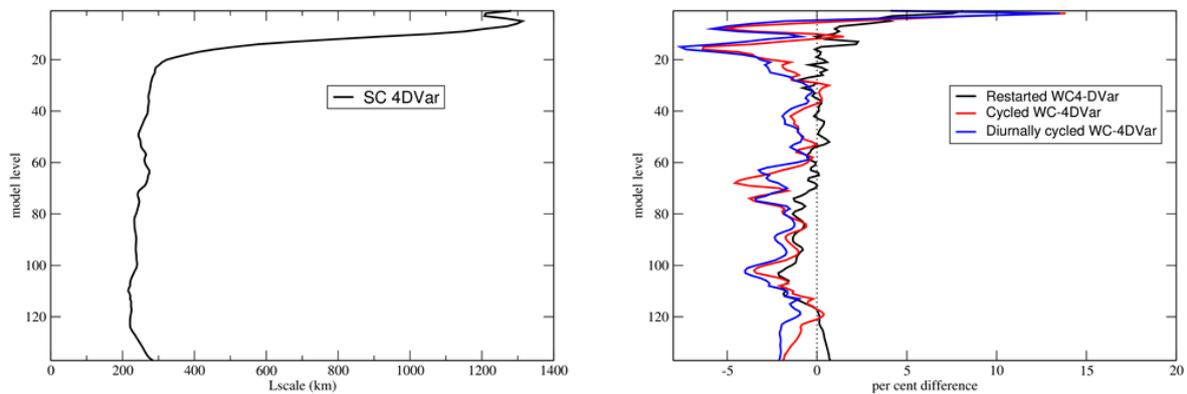

Figure 11: *Left panel: Profile of globally averaged length-scales of temperature analysis increments of the strong constraint 4DVar experiment. Right panel: Profile of the relative difference of the length scales of the restarted WC 4DVar (black), the cycled WC 4DVar (red) and the diurnally cycled WC 4DVar (blue) experiments with respect to those of the SC 4DVar. The values have been computed over August 2019.*

## 5. Data assimilation diagnostics

The effectiveness of a data assimilation system is ultimately determined by its capacity to make the best possible use of the available observations, given the forecast model and what is known about their respective errors. For this reason, we discuss here diagnostics from assimilation experiments of the WC-4DVar setups described before, conducted over a two months period (16 July to 24 September 2019). The baseline is the ECMWF IFS system run in the current (from July 2020) WC-4DVar operational configuration.

In terms of temperature mean errors, the plots in Figure 12 show that all experiments behave similarly in the stratosphere (above 100 hPa) with the exception of the restarted WC-4DVar, which performs worst. This confirms the diagnostic results presented in Sec. 4, i.e. the cycling of model error information is useful in the stratosphere, where diurnal effects are small. In the troposphere, the two experiments which cycle the model error information (i.e., cycled and diurnally cycled WC-4DVar, red and black lines) have a larger impact on the mean errors than the experiments which do not cycle it (restarted WC-4Dvar, green) or which do not apply the model error correction in the troposphere (Operational WC-4DVar, blue). Note how the cycled WC-4DVar experiments are able to drastically reduce the mean background errors for the aircraft observations (top row), while they appear to degrade them for the radiosonde observations near the tropopause and in the boundary layer (but not the bottom layers). One possible explanation is that weak constraint 4DVar works by implicitly debiasing the forecast model against the analysis (Laloyaux et al., 2020b; Eyre, 2016). As the tropospheric temperature analysis in regions covered by the radiosonde network is influenced by aircraft temperature observations, due to their much larger numbers and temporal frequency, this may explain the results shown here. This is particularly true in the boundary layer (1000-850 hPa), while around the tropopause level the influence of other observing systems may also be significant.

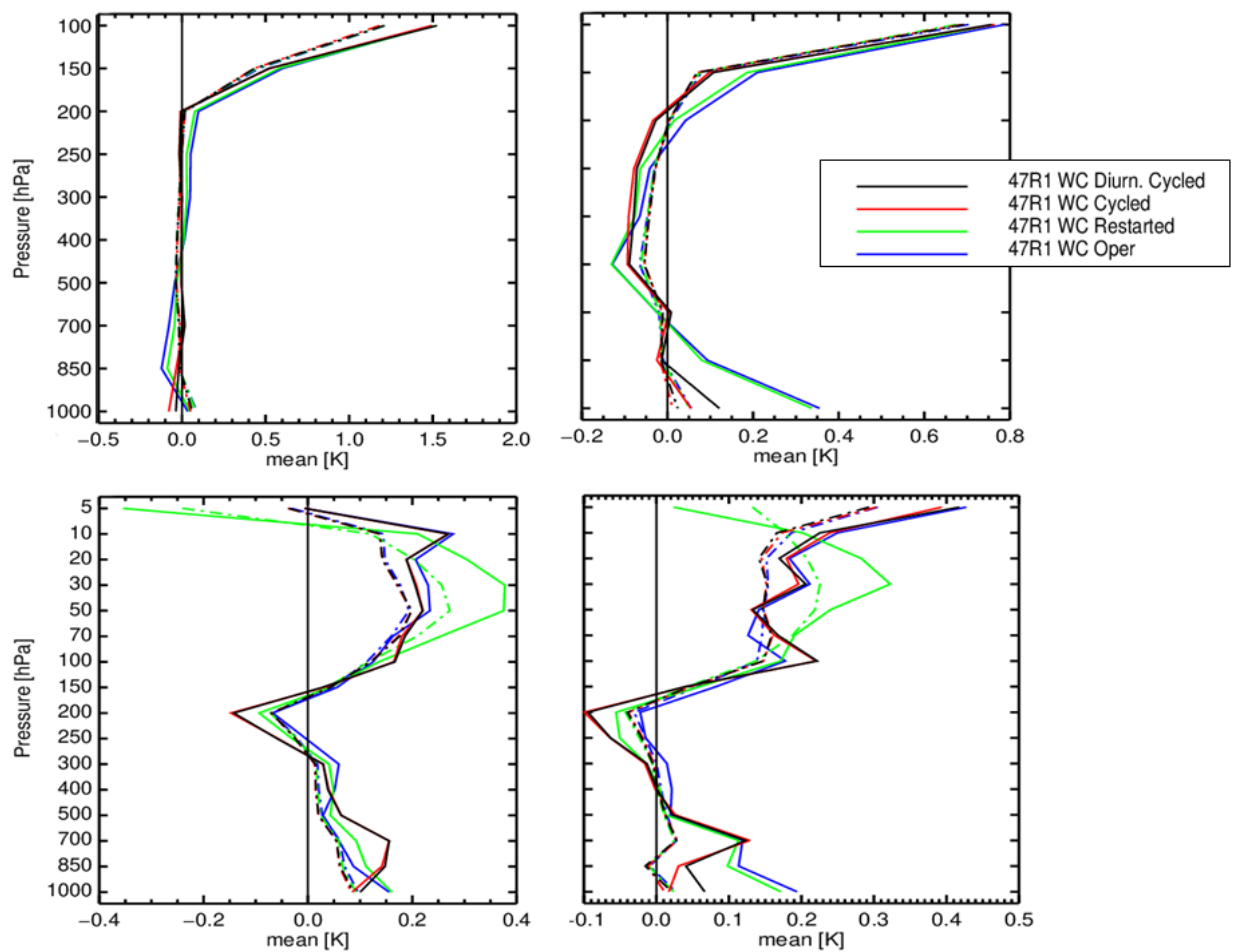

Figure 12 *Top row: Profiles of mean background departures (O-B, continuous lines) and analysis departures (O-A, dashed lines) for aircraft temperature measurements in the northern (left panel) and southern hemisphere (right) for the experiments discussed in the main text. Bottom row: as top row for radiosonde temperature measurements.*

The effect of the tested WC-4DVar configurations on the wind and humidity systematic errors in shown in Figure 13. Again, only the cycled WC-4DVars have a significant effect on the mean errors. This effect is clearly positive for tropospheric winds and it is mainly driven by aircraft wind observations. On the other hand, the effect is negative for boundary layer humidity, even though the degradation is small in absolute and relative terms (the mean errors are about one order of magnitude smaller than the O-B standard deviation). As the humidity model error is not explicitly estimated and corrected in the current version of the ECMWF WC-4DVar, this suggests that it could be useful to extend the current model error control variable to estimate humidity corrections consistently with the temperature corrections.

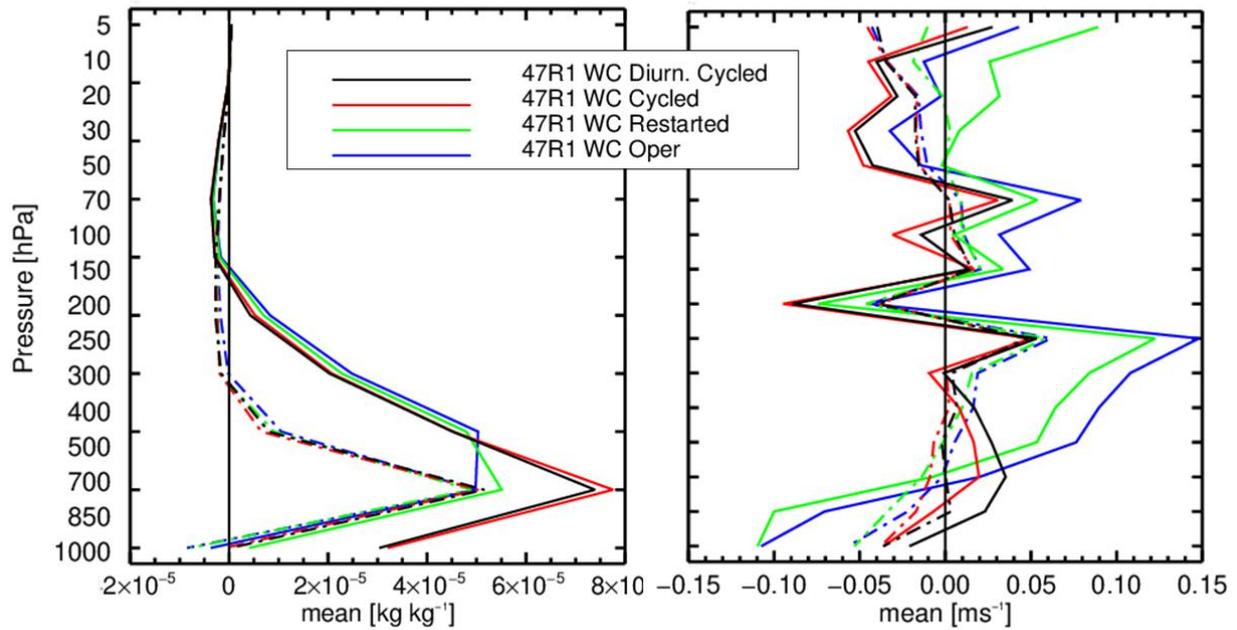

Figure 13 *Profiles of mean background departures (O-B, continuous lines) and analysis departures (O-A, dashed lines) for radiosonde humidity measurements (left panel) and for conventional (radiosonde, aircraft, pilot, profilers) wind observations (right panel) for the experiments discussed in the main text.*

In terms of the random component of initial errors, Figure 14 shows the impacts on some of the main observing systems which measure atmospheric temperature (top row), and wind (bottom row). The analysis departures for all experiments (left panels) are generally smaller than for the operational WC-4DVar. This is to be expected, as all experiments provide additional degrees of freedom for the minimisation to fit observations in the troposphere. The question is whether the model error learned in the minimisation reduces observation departures variability in the successive background forecast (right panels). This appears to be mostly true, particularly in the boundary layer and more clearly for the cycled model error experiments (black and red lines) than for the restarted WC-4DVar experiment (green lines). The exceptions are for the AMSU-A microwave sounder, where the background fit is degraded for the cycled WC-4DVar experiment (red lines) in the tropospheric sounding channels, while the fit of the restarted WC-4DVar experiment (green) is degraded in the stratosphere. The degradation of the restarted 4C-4DVar appear to be a robust signal, as it is confirmed by other temperature-sensitive radiometers (not shown), but it is probably smaller in magnitude than it appears from this plot as this experiment ingests about 2% more AMSU-A observations than the all the others.

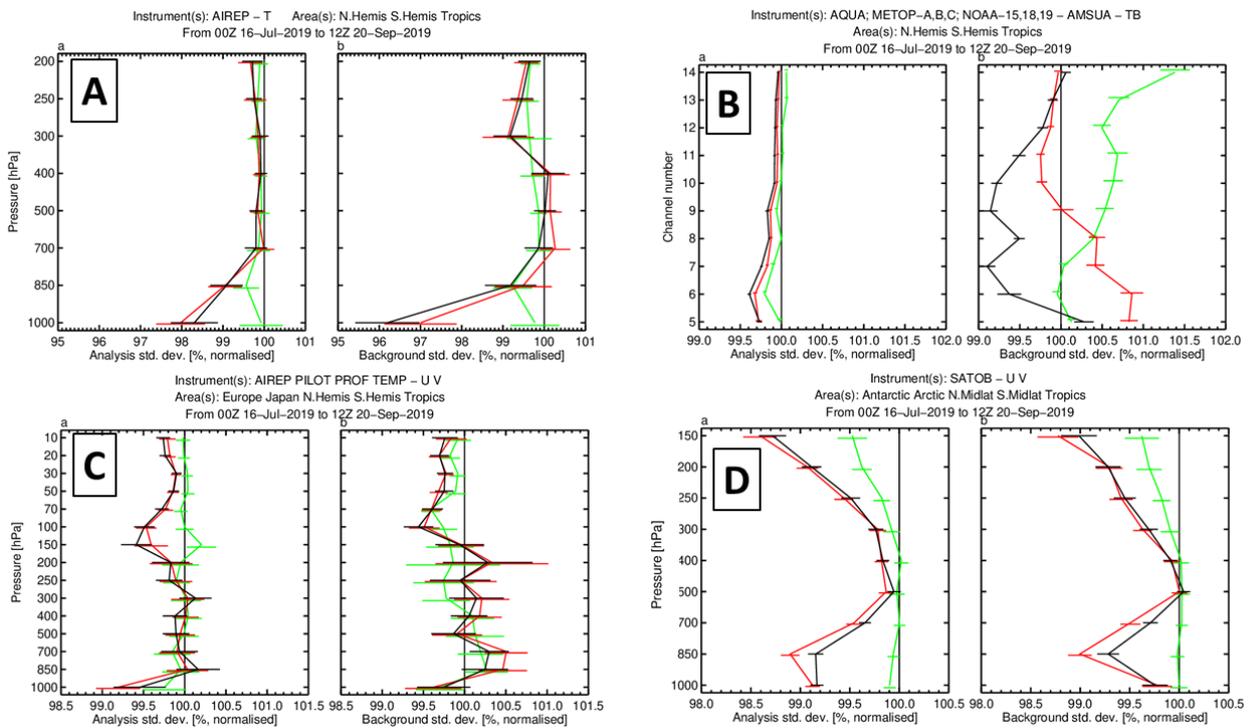

Figure 14: *Profiles of the normalised standard deviation of analysis departures (O-A, left plots), left, and background departures (O-B, right plots) for aircraft temperature observations (panel A); AMSU-A microwave radiances (panel B); conventional wind observations (panel C); and Atmospheric Motion Vectors observations (panel D). The normalisation is done with respect to the operational WC-4DVar. Legend as in Fig. 18. Statistics collected over the 16/07/2019 to 20/09/2019 period.*

In terms of forecast skill measures, an in-depth analysis would require verifying the forecasts accuracy with respect to a set of unbiased observations, such as radiosonde measurements and other conventional observations, as any analysis-based verification will, to some extent, be affected by the systematic errors of the model. This discussion will be deferred to a forthcoming paper. Limiting ourselves to performance measures that are known to be less affected by systematic errors in the verification dataset and more indicative of synoptic performance, like geopotential 500 hPa forecast RMSE, we see (Figure 15) that apart from the first 24-48 hours, the three experiments perform similarly, also with respect to current stratospheric-only WC-4DVar. The fact that all the experiments skill scores show a jump (in either direction), during the first 24-48 hours of the verification is indicative that the forecast is reverting towards the attractor subspace of the model and away from the initial analyses on the 1 to 2 day timescale. This observation may also explain why the impact on forecast skill scores is small with respect to what could be expected from the observation departures' diagnostics: the model error forcings estimated in WC-4DVar are not applied in the long forecast integrations, but only in the short forecasts used to cycle the assimilation system. The impact of applying the model error forcings in the free forecasts will also be discussed in future work.

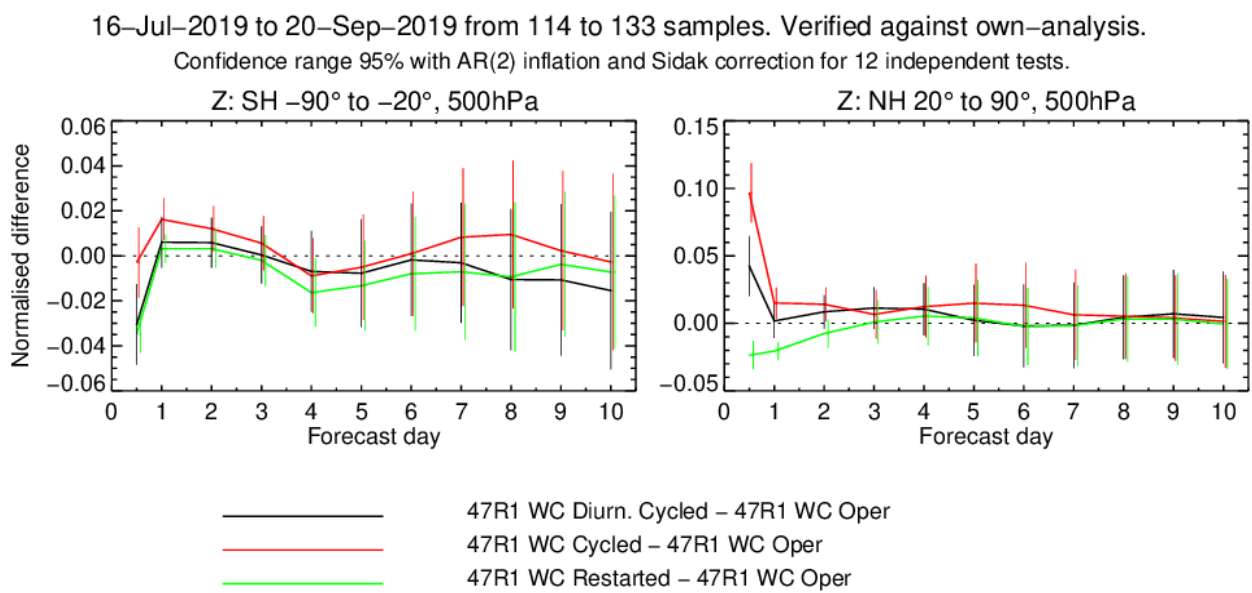

Figure 15: *Normalised RMSE of geopotential forecasts at 500 hPa for the diurnally cycled WC-4DVar (black line), cycled WC-4DVar (red line) and restarted WC-4DVar (green line). The normalisation is done with respect to the operational WC-4DVar. Statistics collected over the 16/07/2019 to 20/09/2019 period.*

## 6. Discussion and Conclusions

Starting from the classical idea of the Lagged Innovation Covariance (Daley, 1992a,b), a similar diagnostic quantity, the Lagged Analysis Increment Covariance (LAIC) can be introduced under a not overly restrictive set of assumptions. We have used LAIC to explore the characteristics of time-correlated model errors in the operational ECMWF analysis system. As noted by Daley (1992a), lagged covariance diagnostics are four dimensional performance measures, and thus affected by all the complexities and imperfections of a real world assimilation system, which makes it difficult to provide an unambiguous interpretation of the results. We have used this diagnostic to explore the evolution of time-correlated model errors under the assumptions that: a) the contribution to LAIC from suboptimalities in the specification of the error covariances is small; and b) the systematic errors affecting observations have been completely or largely reduced through appropriate observation bias correction procedures. To some extent, the validity of both assumptions can be questioned, and this offers scope for further refinements in both the theory and the practical use of these diagnostics.

With the caveats discussed above, we believe LAIC can be a useful additional tool to measure the performance of a data assimilation system and highlight promising directions for its further development. From the results shown in Sect. 3, LAIC confirms the presence of systematic, correlated model errors in the ECMWF assimilation cycle, with a significant diurnal component more evident in the lower troposphere. This result suggests that the current cycling strategy of persisting the model error estimates in time from one assimilation update to the next will need to be revisited. To start exploring this idea, we have run three assimilation experiments using different setups for the cycling of the model error estimates, going from no cycling (restarted WC-4DVar, $\boldsymbol{\eta}^b = \boldsymbol{0}$), to standard cycling (cycled WC-4DVar, $\boldsymbol{\eta}^b = \boldsymbol{\eta}^a_{n-1}$), to diurnal cycling (Diurnally cycled WC-4DVar, $\boldsymbol{\eta}^b = \boldsymbol{\eta}^a_{n-2}$). All these WC-4DVar configurations were shown to be able to partially reduce the effects of model systematic errors in the assimilation cycle in terms of reducing both the magnitude and the time correlations, with the diurnally cycled version more effective than the other two. These results were confirmed by the standard data assimilation diagnostics shown in Sec. 5, where the various WC-4DVar experiments generally showed a closer fit to observations in terms of both mean and variability. We have not provided here an in-depth look at forecast performance, but early indications are that this is generally positive but small. There are indications that this rather surprising result appears to be caused by the tendency of the free-running forecast to quickly revert back to the model (biased) attractor, thus limiting the benefit of the WC-4DVar analysis to forecast performance. This interesting hypothesis will be validated in future research.

From a development perspective, the diagnostic work described in this manuscript has highlighted the importance of some specific research directions. In terms of WC-4DVar development, the current assumption of constant model error forcing during the 12 hour assimilation window is hard to justify in the troposphere and more sophisticated model error models will be required in order to provide an improved representation of the time-dependent aspects of model error evolution. Furthermore, in the current version of WC-4DVar the model error control vector does not contain a humidity variable. This is based on earlier findings that no significant large scale biases were present in the short range humidity forecasts, which was confirmed in more recent studies (Bonavita and Laloyaux, 2020). Background departures for humidity sensitive instruments are generally improved (both in the mean and the standard deviation) in the WC-4DVar experiments, with the notable exception of conventional humidity observations (from radiosonde and aircraft) in the lower troposphere. Whether this is a robust signal that points to the need of an explicit modelling of humidity model errors consistent with the temperature model error corrections, or rather an indication of suboptimalities in the current usage of conventional humidity observations, is the subject of ongoing investigations.

Another way to deal with systematic model errors is through the use of Artificial Neural Networks (ANN), either on their own or through hybrid ANN-WC4DVar configurations (Bonavita and Laloyaux, 2020). The results presented in this work indicate at least two interesting avenues for further development. One concerns the training phase of the ANN, where it might prove useful to increase the temporal frequency of the training dataset (i.e., train the ANN on 6-hourly or 3-hourly analysis increments instead of the current 12-hourly samples) to let the ANN better capture the daily evolution of model errors. The other aspect regards the use of the ANN to correct model errors in a free running model forecast. The presence of diurnal (and possibly higher frequency) model error components makes it essential that the ANN provides online tendency corrections to the model evolution, which in turn would be facilitated by a close integration of the ANN inside the ECMWF IFS. Both these ideas are being pursued and results will be reported in future work. Finally, the LAIC diagnostic itself can be further developed, in particular with regards to exploring its sensitivities to other suboptimalities of the assimilation system (e.g., misspecifications of error covariances, residual biases in observations), preferably in a simplified, controlled testing environment. Other aspects of the LAIC diagnostic also merit further investigation. For example, in this work we have only drawn qualitative conclusions on model error based on the shape of the LAIC diagnostic. It will be interesting to understand under which conditions it may be possible to infer more quantitative results from, e.g. from the temporal decay of the LAIC as a function of increasing lag time.